\documentclass{iopart}
\usepackage{latexsym}

\begin{document}

 \jl{6}
 
\title[QNM level spacing]{Quasi normal modes: A simple derivation of the level spacing of the frequencies
} 
\author{T. Padmanabhan \footnote[1]{E-mail address:  {\tt nabhan@iucaa.ernet.in}}}
\address{Inter-University Centre for Astronomy and Astrophysics, Post Bag 4, Ganeshkhind, Pune - 411 007.}

\begin{abstract}
It is known that the imaginary parts of the frequencies of the quasi normal modes of the
Schwarzschild black hole
are equally spaced, with the level spacing dependent only on the surface gravity.  We generalize this result  to a wider class of spacetimes and provide a simple derivation of the imaginary parts of the frequencies.
The analysis shows that the result is closely linked to the thermal nature of horizons and arises from the
exponential redshift of the wave modes close to the horizon.
\end{abstract}

  In the study of small perturbations around a black hole spacetimes, quasi-normal modes (QNM, hereafter)
  play a crucial role and have been investigated very extensively (for a review, see \cite{qnmreview}).
  It is known that the frequencies of QNM for the Schwarzschild black hole have the structure 
  \begin{equation}
  k_n =   i\kappa\left(n+\frac{1}{2}\right)+\frac{\ln 3}{2\pi} \, \kappa  + \mathcal{O}[n^{-1/2}] 
\end{equation}
  where $\kappa$ is the surface gravity of the black hole. This result was originally obtained
  numerically \cite{nollert,andersson} 
  and, surprisingly enough, was given an analytic proof only recently \cite{motl,MN}.
  Most of the theoretical studies were either model dependent or concentrated on the 
  real part of $k_n$ because of its relationship to loop quantum gravity  and the
  area spectrum of black holes (see, for an incomplete  sample of references,
   \cite{hod,dreyer,Kun,oppen}).
  It is, however,  equally intriguing why the imaginary part of $k_n$ has a, simple, equally spaced 
  structure with the surface gravity of the horizon determining the level spacing. We shall try to provide a simple
  derivation of this feature in a context more general than that of a Schwarzschild black hole in this paper.
  [While this work was nearing completion, a paper \cite{mmv}
    appeared in the arXive with substantially same results. We shall provide brief comparisons wherever relevant.]
    
    For our analysis, we shall choose a general class of spherically symmetric metrics
    of the form 
    \begin{equation}
     ds^2=-f(r)dt^2+f(r)^{-1}dr^2 + r^2(d\theta^2+\sin^2\theta
     d\phi^2)
     \end{equation}
   It was shown in \cite{tpcqg}
   that spacetimes described by the above class of metrics
have a fairly straight forward thermodynamic interpretation and --- in fact --- Einstein's equations
can be expressed in the form of a thermodynamic relation $TdS=dE-PdV$ for such spacetimes. Since quasi normal modes are conjectured to have possible, though as yet unclear,  relationship with horizon thermodynamics,
it will be interesting to explore the QNM for these metrics. We shall show, by a fairly straight forward 
and simple analysis that the imaginary part of the QNM frequencies for these spacetimes is given by   Im $ k_n=in\kappa$ for
sufficiently large $n$. Our analysis shows that this result has the same mathematical origin as the standard
Hawking evaporation, in the sense that both arise due to similar mathematical operations involving infinite redshift surface and tortoise coordinate.  What is more, we shall only use the behaviour of $f(r)$ on one side of the horizon; for example, in the case of Schwarzschild black hole we will obtain the results sticking to the $r\geq 2M$ region, unlike for e.g., the results in ref. \cite{MN} which depends on the form of the metric near $r=0$.

   We are interested in the case in which the spacetime has one (or more) horizon(s)
   at which $f(r)=0$ and $f'(r) $ is a finite constant. (The second condition rules out
   extremal black holes and similar spacetimes.) The surface gravity is then given by
   $\kappa = (1/2)|f'(a)|$ where $a$ is the radius of the horizon. For simplicity, we shall consider
   the single horizon situation for most of the analysis and will comment on generalization for 
   multi horizon scenarios (which turns out to be far from  trivial) in the end.
   It was  shown in \cite{tpcqg}
   that the Einstein's equations allow us to write $f(r)$ in the form 
   \begin{equation}
   f(r) = 1 - \frac{a}{r} - \frac{1}{r} \int_a^r \rho(r) r^2 dr
   \end{equation}
   where $\rho(r)$ is the energy density and we have chosen the normalization such that
   $f=0$ at the $r=a$ surface which acts as a horizon. This provides an explicit context for 
   our discussion. 
   
   Consider a massless scalar field $\phi$ satisfying the wave equation $\Box \phi =0$ in this spacetime.
   (Generalization to higher spins are straight forward).
   We look for solutions to the wave equation in the form
   \begin{equation}
\phi= {1\over r} \, F(r) \, Y_{\ell m}(\theta,\phi)\,
\exp(ikt) \, ,
\end{equation}
   Straight forward algebra now leads to a ``Schrodinger equation'' for $F$ given by:
   \begin{equation}
   \left[-\frac{d^2}{dr_*^2} + V(r)\right] F =k^2 F
   \label{seqn}
   \end{equation}
   where
   \begin{equation}
   V(r) =f\left( \frac{l(l+1)}{r^2} + \frac{f'}{r}\right) ; \qquad f' \equiv \frac{df}{dr}
   \label{vofr}
   \end{equation}
   and $r_*$ is the tortoise coordinate defined by 
   \begin{equation}
   r_* = \int \frac{dr}{f(r)}
   \label{rstar}
   \end{equation}
   In equation (\ref{seqn}) we treat $r_*$ as the independent variable and all other quantities
   are expressed in terms of $r_*$ through the function $r_*(r)$. 
   Near the horizon $r=a$, we can expand $f(r)$ in a Taylor series and use (\ref{rstar})
   to obtain an explicit relationship between $r$ and $r_*$: 
   \begin{equation}
   r_* = \frac{1}{2\kappa} \ln \left( \frac{r}{a} -1 \right) +  L\left( \frac{r}{a} -1 \right) + \mathcal{O}[(r-a)^2]
   \end{equation}
   where the length scale $L$ depends on the source distribution $\rho(r)$.
   [At a fundamental level, so does the surface gravity $\kappa$; however, we shall see that
   our results are independent of $L$ but depends on $\kappa$.] With this definition, $r\ge a $ is mapped
   to $r_*\ge -\infty$, and we will not require the region $(r<a)$ beyond the horizon.
   
   We shall attempt to determine the frequencies of QNM by treating equation (\ref{seqn})
   as a one-dimensional Schrodinger equation. In that case, the frequencies of QNM
   can be identified with the poles of the scattering amplitude $S(k)$ in the momentum space. The scattering
   amplitude, in turn, requires a complete solution of equation (\ref{seqn}) which, of course, is not
   available. We shall therefore invoke two approximations to obtain the scattering amplitude.
   First, we shall invoke the Born approximation for $S(k)$,  which should be valid for large
   momentum transfer and hence for $n\gg 1$ limit of QNM. Second, we shall evaluate
   the integral in Born approximation by taking the leading order contribution near the horizon.
   We shall first obtain the result and then comment on the validity of the approximations.
   
   The scattering amplitude in the Born approximation is given by the Fourier transform
   of the potential $V({\bf x})$ with respect to the momentum transfer ${\bf q} = {\bf k}_f -{\bf k}_i$.
   In one dimension, ${\bf k}_i$ and ${\bf k}_f$ should be parallel or anti-parallel; further we can take their magnitudes to be the same for scattering in a fixed potential. Then non-trivial momentum transfer occurs only for ${\bf k}_f = - {\bf k}_i$ so that $q = 2k_i$ in magnitude.
    As we shall see, the factor 2 in this relation is crucial to get the correct result and it arises due to the
    constraint  ${\bf k}_i + {\bf k}_f =0$. (Our original problem was three-dimensional and we are {\it not}
    working out the three-dimensional scattering amplitude in, say, $s$-wave limit. Rather, we first map the problem to an one-dimensional Schrodinger equation and study the scattering amplitude in one dimension). 
    The scattering amplitude can now be expressed as 
    \begin{equation}
    S(k) = \int_{-\infty}^\infty dr_* V(r_*)\, e^{2ikr_*}
    \end{equation}
 where we have omitted irrelevant constant factors.   Changing the integration variable to $r$ using (\ref{rstar}) and the form of the potential in (\ref{vofr}),
    we get 
    \begin{equation}
    S(k) = \int_0^\infty \frac{d\epsilon}{a} \left[ \frac{l(l+1)}{(1+\epsilon)^2}+ \frac{af'}{(1+\epsilon)}\right] \epsilon^{ik/\kappa} e^{i2kL\epsilon}
    \label{keyint}
    \end{equation}
    where $\epsilon \equiv [(r/a)-1]$. This integral, unfortunately, cannot be expressed in closed form
    (except in a few special cases; see equation (\ref{exact}) below).
    However, it is obvious that the dominant  contribution, which is of interest to us, arises close to the horizon near $\epsilon  \approx 0$. (This becomes clearer if we rotate
    the contour of integration to purely imaginary values of $\epsilon$ by $\epsilon\to i\epsilon$ so that the integrand takes the form $Q(x)x^{i\alpha}e^{-x}$ with $Q$ being regular near $x=0$). Since $f'$ is regular at $\epsilon=0$, the leading order contribution near
    $\epsilon =0$ has the structure
     \begin{equation}
    S(k) \propto \int_0^\infty \frac{d\epsilon}{a} \epsilon^{ik/\kappa} e^{i2kL\epsilon} \propto
          \Gamma\left(1+ i\frac{k}{\kappa}\right)
          \label{approxint}
    \end{equation}
    where we have ignored a finite proportionality constant which is irrelevant for determining
    the poles. The poles of the scattering amplitude occur when the argument of the gamma function
    take negative integral values. This immediately gives the imaginary part of the QNM frequencies
    to be 
    \begin{equation}
    k_n  = in \kappa \qquad ({\rm for}\ n\gg 1)
    \label{poles}
    \end{equation}
    This completes our derivation and we shall now comment on several features of the 
    analysis.

     To begin with, let us note that
    there is a simple connection between the above result and the standard derivation of
    Bogoliubov coefficients in the case of a horizon which causes infinite redshift.
    A wave mode, the frequency of which is being exponentially redshifted,
    will be described by the amplitude of the form $\phi(t) = \exp(iqe^{-\kappa t}) $ where $q$
    is a constant. Such a form arises in several contexts like the black hole spacetime, Rindler
    frame and in  a  universe which is asymptotically De Sitter. An observer using the time coordinate
    $t$ will decompose this mode into positive and negative frequencies with respect to $t$ by
    the relation
    \begin{equation}
  \phi(t) = \int_{-\infty}^\infty \frac{d\nu}{2\pi} f(\nu) e^{-i\nu t}
  \label{ftf}
  \end{equation}
    Elementary calculation shows that $f(\nu)$ has the factors $\Gamma(-i\nu/\kappa)$ and
    $\Gamma(1-i\nu/\kappa)$ for positive and negative  values of $\nu$; hence $f(\nu)$ has 
     poles exactly as in  (\ref{poles}).
    The integrals
    involved in this problem are identical to those we encountered above to obtain the scattering
    amplitude. The gamma function arises in both the cases due to the logarithmic behaviour of 
    the tortoise coordinate or, equivalently, the exponential behaviour of $r$ as a function of
    $r_*$ near the horizon.  Thus, the structure of imaginary part of the QNM frequencies is closely linked to the 
    standard analysis performed for deriving the thermal nature of horizons.
    
    In the case of Schwarzschild (or De Sitter spacetime), the integral in (\ref{keyint}) can be 
    evaluated in closed form in terms of confluent hyper-geometric function ${}_1F_1$  using the
    result: 
    \begin{eqnarray}
    I &=& \int_0^\infty dx \frac{x^ae^{-x}}{(1+bx)^n} = b^{-n} \Gamma(1+a-n) \, {}_1F_1(n,-a+n,b^{-1})\nonumber\\
    &&+ \frac{b^{-(1+a)} \Gamma(1+a) \Gamma(-1-a+n)}{\Gamma(n)}\, {}_1F_1(1+a,2+a-n,b^{-1})
    \end{eqnarray}
    In the case of Schwarzschild spacetime, $f=1-(a/r)$ and the scattering amplitude (omitting
    unimportant constant prefactors) becomes with $\theta\equiv k/\kappa$,
    \begin{eqnarray}
      \label{exact}
    S(k) &=& \left(\frac{i}{2kL}\right)^{2-i\theta} | \Gamma(1 +i\theta)|^2 
   \Big[ l(l+1) \, {}_1F_1(1+i\theta, i\theta, -2ikL)\nonumber\\
   &&\qquad\qquad+\frac{1}{2} (1-i\theta)  \, {}_1F_1(1+i\theta,i\theta -1,-2ikL)\Big]\nonumber\\
    && +\Gamma(-2+i\theta) \Big[ {}_1F_1(3,3-i\theta,-2ikL)\nonumber\\
    &&\qquad\qquad- \frac{l(l+1)}{2kL}(2i +\theta)
     \, {}_1F_1(2,2-i\theta,  -2ikL)\Big]
    \end{eqnarray}
    This exact expression confirms the structure of the poles identified earlier; the Gamma function
    has poles when its argument takes negative integral values and ${}_1F_1(x,y,z)$ has poles
    when $y$ takes negative integral values or zero. All these poles are summarized
    by equation (\ref{poles}) for large $n$. It is clear that no new poles appear at least in the
    case of Schwarzschild geometry when the exact analysis is performed. 
    
     The integral in (\ref{keyint}) can also  be evaluated by retaining the full Taylor
    series expansion in $\epsilon$ as is done in ref.\cite{mmv}.
    Since each term of the power series is a gamma function, our key result is not changed.
    However, it is not clear whether such an expansion, retaining all the powers of $\epsilon$, 
    is consistent with the Born approximation. It seems more natural to relate the result 
    to the behaviour near the horizon which is what we have done. This is particularly true if the 
    spacetime has a second horizon at $r=b$, say, in which case it is probably incorrect
    to extend the range of integration to all values of $r$. 
    There is also the issue of new poles appearing when an infinite series is summed up which 
    are not present in the sum containing finite number of terms (like the pole at $x=1$ which arises
    when the series $(1+x + x^2+\cdots)$ is summed up retaining all the terms). This does not happen in the case of Schwarzschild 
    geometry, as is evident from the exact result in equation (\ref{exact}) and is unlikely to happen
    in a more general situation. We therefore can conclude that the poles of the scattering
    amplitude \emph{in the Born approximation} occur  along the imaginary axis.

     While the analysis appears to be strikingly simple, there are some subtleties which are not
    fully resolved. Since the result arises from the nature of the potential near the 
    horizon, it seems reasonable to use an approximate form of the potential valid near the 
    horizon in the equation (\ref{seqn}) and solve for the scattering amplitude. It is fairly straightforward
    to do this (see,e.g., ref.\cite{solo}).
    To the lowest order, the horizon geometry can be approximated by that of a Rindler spacetime
    and the solutions to equation (\ref{seqn}) can be expressed in terms of the Macdonald function
    $K_{i\nu}(z)$. The scattering amplitude has unit modulus but exhibits divergent phase on the 
    poles determined by equation (\ref{poles}). At this order, only the surface gravity of the horizon
    plays any role and the analysis is equivalent to what we have done.  Mathematically, this corresponds to 
    retaining the lowest order contribution in the term in the square brackets in equation (\ref{keyint}).
    The next order approximation to the potential near the horizon is provided by retaining one higher order in
    $\epsilon$ in evaluating the integral in (\ref{keyint}) which will lead to the sum of two Gamma functions
    in our approach, without changing the structure of the poles. However, a different picture emerges if we solve the Schrodinger equation {\it exactly} in this {\it approximate} potential.
    When the potential is approximated by retaining one order higher than the 
    Rindler case, then a more complicated form of scattering amplitude results (see equation 11 
     of ref.\cite{solo}).
    The imaginary part of the QNM frequencies obtained by this method has a 
    level spacing which is $2\kappa$ rather than the standard result $\kappa$ (see equation 12
     of ref.\cite{solo}; this is 
    possibly related to the difference between $q=2k_i$ and $k_i$ in the case of Born
    approximation mentioned earlier).  
    It is not clear whether {\it approximating} the potential, to a finite order of accuracy near the 
    horizon and then solving the Schrodinger equation {\it exactly} in the resulting potential, is a 
    consistent procedure. In general, it is difficult to compare the order of accuracy of different
    approaches and further work is required in this direction.

    Our analysis uses a metric which covers not only the Schwarzschild spacetime but
    also several generalizations including Reissner-Nordstrom, De Sitter and Schwarzschild-De Sitter.
     While the naive integration performed in (\ref{approxint}) uses only the form of the 
    metric near the horizon and will go through for all the cases, there are other subtleties
    which must be borne in mind while generalizing this result. In the case of De Sitter spacetime,
    for example, the lack of asymptotic flatness will be an issue in defining the QNM in our approach.
    In SdS, the situation is much worse since there is no definition of global temperature for the
    spacetime with more than one horizon. Naive application of (\ref{approxint}) will now pick
    up contributions from the two horizons additively if we suitably restrict the range of integration 
    in equation (\ref{keyint}).
    It is not obvious that multi horizon spacetime can be handled in such a manner and this issue
    is under investigation \cite{trc-tp}.
    
    Finally, there are (at least!) two questions of conceptual nature which we have not answered. Physical quantities with a quantized spectrum are of special interest when they have constant spacing. In the case of horizon area, for example, one can attempt to relate this to the intrinsic limitations in  measuring length scales smaller than Planck Length
\cite{tplimit}. But the uniform spacing of QNM frequencies is a purely {\it classical} result and hence is harder to understand physically. While  relating it to the poles of the Gamma function could be a step forward, we have a long way to go in understanding it. Second, the procedure does not provide an insight into the real part of $k_n$ and it does not seem easy to relate this result to previous analytic derivations of Re $k_n$.
The results in ref. \cite{MN}, for example, depends crucially on the form of the metric near $r=0$,
while we have only used the form of metric in the range $r\ge a$.
      
I thank T. Roy Choudhury for comments on the draft of the paper.

\end{document}